\documentclass[twoside,slac_one]{revtex4}
\usepackage{graphicx}
\usepackage{fancyhdr}
\usepackage{amsmath} 
\usepackage{bm}
\usepackage{amsxtra}
\usepackage{amssymb}
\usepackage{amsthm}
\usepackage{latexsym}
\usepackage{lscape}
\usepackage{epsfig}
\usepackage{epstopdf}
\DeclareGraphicsExtensions{.pdf,.eps,.png,.jpg,.mps}

\begin{document}

\title{Belle II Detector: Status and Proposed US Contributions}

%

\author{J. Fast}
\affiliation{Pacific Northwest National Laboratory, Richland, WA, USA \\ for the Belle II Collaboration}

\begin{abstract}
High precision flavor physics measurements are an essential complement to the direct searches for new physics at the LHC. Such measurements will be performed using the upgraded Belle II detector and upgraded KEKB accelerator. The status of the Belle II detector and proposed role of the US Belle II collaborators are presented in this article.
\end{abstract}

\maketitle

\thispagestyle{fancy}


\section{Introduction}
The B factory experiments, Belle at the KEKB collider at KEK~\cite{belle} and BaBar at the PEP II collider at SLAC~\cite{babar}, were built to measure the large mixing-induced CP violation in the B0 system predicted by the theory of Kobayashi and Maskawa~\cite{km}. The successful confirmation of the prediction led to the Nobel Prize for both theorists.

The B factories were built to answer the question Is the CKM description in the Standard Model correct?
Most B factory results are in good agreement with the expectations from the Standard Model (SM) and confirm the CKM structure of quark mixing and CP violation, but some measurements show tensions with the SM prediction. The Super B factories will address the question In what way is the Standard Model wrong?
Much larger datasets are needed for high-precision measurements to search for significant deviations from the SM which are expected to exist.  The SuperKEKB and Belle II detector are being constructed to perform these high precision studies with a dataset $50 ab^{-1}$.

The SuperKEKB accelerator will operate at an instantaneous luminosity of $8 \times 10^{35} cm^{-2}s^{-1}$, a factor of 40 times higher than the luminosity of the KEKB accelerator at the end of the Belle physics program.  The increased luminosity will derive primarily from shrinking the beam size at the interaction point (nano beams) utilizing new final focus quadrapole magnets in combination with a factor of ~2 increase in beam currents.  To counteract the increase in the beam-beam parameter due to the reduced beta functions, the beam emittance must also be reduced; this will be achieved with a new electron source and a new damping ring.

\section{Belle II Detector}
\begin{figure*}[htb]
\centering
\includegraphics[width=135mm]{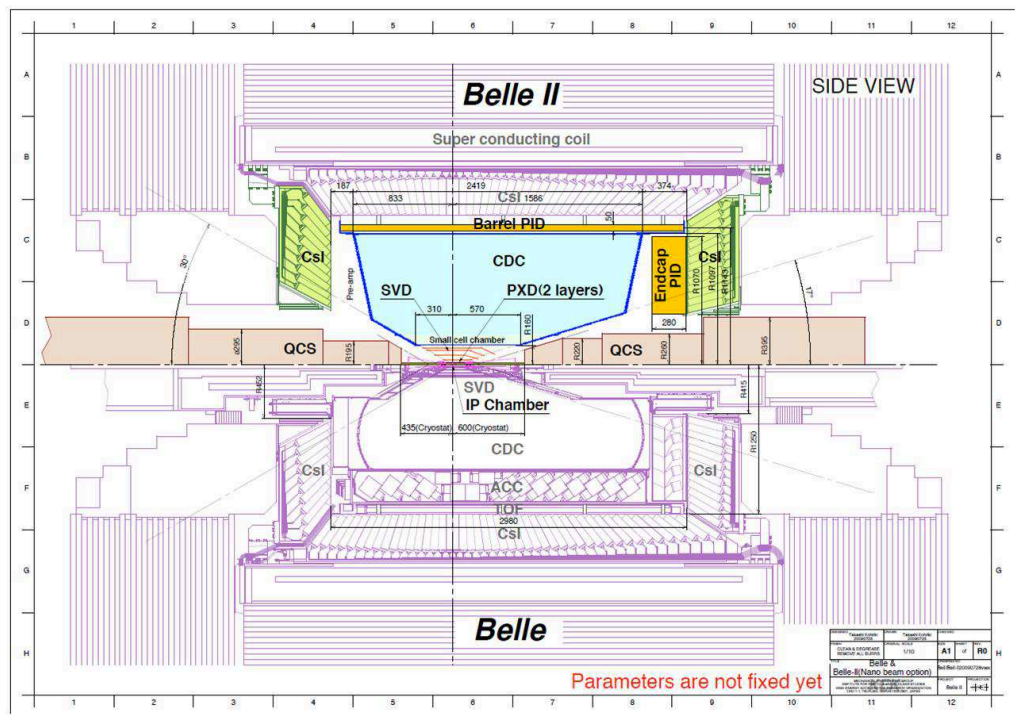}
\caption{The Belle II detector (top half) compared with the Belle detector (bottom half).}
\label{fig_detector}
\end{figure*}

The high luminosity of the Super KEKB accelerator will produce higher rates and radiation levels in the Belle detector.  The Belle II detector upgrade~\cite{tdr} will address these challenges as well as improving tracking performance and particle identification capability. In addition the increased event rate puts high demands on trigger, data acquisition, and computing. To cope with the conditions at the SuperKEKB collider, the tracking and particle identification systems of the Belle detector will either be upgraded or replaced by new ones. Figure~\ref{fig_detector} shows a comparison of the Belle and Belle II detectors.

\begin{figure*}[ht]
\centering
\includegraphics[width=135mm]{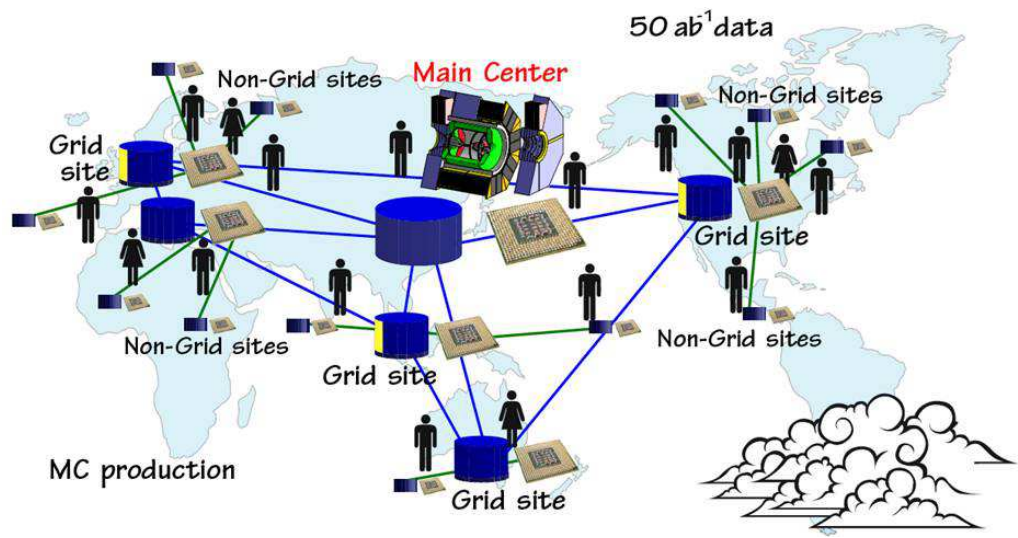}
\caption{The Belle II distributed computing model.}
\label{fig_computing}
\end{figure*}

The innermost part of the tracking system consists of two layers of silicon pixel sensors (PXD) based on DEpleted P-channel Field Effect Transistor (DEPFET) technology. The excellent impact parameter resolution in the beam direction, $\approx 20 \mu m$,  of the PXD will provide improved determination of the vertex position. The PXD is surrounded by four layers of double sided silicon strip detectors (SVD). The larger outer radius of the SVD compared to Belle gives an increase in efficiency of about 30\% for the reconstruction of $K_S \rightarrow \pi^+\pi^-$ decays inside the SVD. A precise measurement of the momentum of charged tracks is provided by the central drift chamber (CDC). Improvements in the momentum resolution compared to the Belle CDC are achieved by a larger outer radius and a smaller cell size.
For the identification of charged hadrons, the time-of-flight detector at Belle will be replaced by an imaging time-of-propagation counter (iTOP). The usage of timing information of internally reflected Cherenkov light allows for a compact design of this particle identification device in the barrel region. The forward region will be instrumented with new aerogel RICH detector system (ARICH) using  layers with different refractive index to generate Cherenkov rings with the same radius for each layer. A kaon identification efficiency of $>\sim 99\% (96\%)$ at a pion mis-identification rate of $<\sim 0.5\% (1\%)$ is expected for $B \rightarrow \rho\gamma$ events reconstructed in the iTOP counter (4~GeV particles reconstructed in the ARICH).
The crystals of the Belle electromagnetic calorimeter (ECL) will be reused for Belle II. To improve the signal to background separation under the higher background conditions at SuperKEKB, the electronics will be upgraded to enable wave form sampling. Muons and $K_L$ mesons are identified by resistive plate chambers (RPC) in the outer part of the Belle detector (KLM). For Belle II the endcap regions and the inner two layers of the barrel region will be upgraded with scintillator strips to cope with the higher background rates and to alleviate concerns about the longevity of the RPC detectors in the higher radiation field regions.  The plastic layers in the barrel also shield the remaining RPC layers from neutron backgrounds.
The almost two orders of magnitude higher rate of interesting physics events requires an upgrade of the data acquisition system and the offline computing system.  In contrast to the KEK-centric computing model of Belle, the Monte Carlo production and physics analysis at Belle II will be done in a distributed way exploiting grid and cloud technologies as shown in Figure~\ref{fig_computing}.

\section{Status and Proposed US Role in Belle II}
The Belle II collaboration was formed in December 2008 and the majority of the US institutions on Belle became founding members of the collaboration.  Since that time the collaboration has grown to ~400 scientists from 57 institutions in 13 countries.
In January 2010 the Japanese government approved the SuperKEKB upgrade and began funding the project with an initial allocation of 100 oku-yen (~\$110 M US).  Subsequently the Japanese government approved the full budgetary request for the upgrades of both the accelerator and detector.  In addition to this funding, significant in-kind contributions will come from the international collaborators, including the US.

The US membership has expanded over the past two years to 42 members for 7 University groups and one national laboratory.  The US groups propose to make significant contributions to the particle identification systems, the electronics upgrades across many of the detector systems, beamstrahlung
monitors for the SuperKEKB accelerator, and hosting a Tier 1 computing facility for Belle II analysis.
For the iTOP system, the proposed US responsibilities include specification and procurement of the fused silica optical elements, multi-anode photomultiplier tube readout electronics including custom application specific integrated circuits (ASIC), firmware and software development, and integration and commissioning activities at KEK.  The US groups provided the current RPC detectors for the KLM system and are proposing to provide the replacement scintillator panels for the inner layers of the barrel as well as the electronics upgrade for the complete endcap and barrel KLM system, including the RPC detectors that will remain in the barrel.  As with the iTOP system, the US groups will provide on-site support during installation and commissioning of the KLM upgrades.  In addition to the custom ASIC-based electronics for the iTOP and KLM, similar readout for the ECAL upgrade and beamstrrahlung monitors will be developed and provided.  This detector upgrade project has been proposed to the US DOE.  Currently the project has CD-0, Mission Need, in the DOE project approval process.  In addition to this set of detector upgrade activities, the US groups have developed beamstrrahlung monitors which are essential to tuning the beam collisions in the nano- beam scheme.  Finally, a Tier 1 computing facility will be hosted by the Pacific Northwest National Laboratory.  In the wake of the Fukushima earthquake and tsunami the US has stood up a computing environment equal to the Belle computing system at KEK to relieve pressure on availability of that system.  This represents a ~10\% test system for the proposed Belle II Tier 1 system.

\bigskip 

\end{document}